# The K-User Interference Channel: Strong Interference Regime

Reza K. Farsani[1]
Email: reza_khosravi@alum.sharif.ir

***Abstract:*** **This paper gives a solution to one of the long-standing open problems in network information theory: "*What is the generalization of the strong interference regime to the K-user interference channel?*"**

***Index Terms:*** **multi-user interference channels, strong interference regime.**

## I. INTRODUCTION

One of fundamental open problems in network information theory is to determine the capacity region of the Classical Interference Channel (CIC). The importance of this problem is by now widely acknowledged. As these channels are very useful models for wireless communication systems, in recent years they have been extensively studied. For a detailed review of the existing literatures refer to Part I of our multi-part paper [1]. However, for the two-user CIC capacity results are known in some special cases, the multi-user channels are far less understood [9, page 6-64].

In 1981 [10], Sato derived a regime for the two-user Gaussian CIC wherein joint decoding both messages at both receivers is optimal and achieves the capacity. Six years later, in 1987 [11], Costa and El Gamal extended the Sato's result for the discrete channel. This regime in which the capacity region is derived by decoding both messages at both receivers is called the "strong interference regime". From that time, for about 25 years, it has been an open problem [9, page 6-68] that what is the generalization of the strong interference regime for the multi-user CICs? In this paper, we give a solution to this problem. Clearly, we develop a new approach based on which one can derive strong interference regime for any given interference network. To this end, we develop some new technical lemmas which have a central role in our derivations.

This manuscript addresses our main results for the classical interference channel. However, our methodology is applicable for arbitrary single-hop communication networks with any topology. Please refer to Part III of our multi-part paper [3] where we have presented a general formula to derive strong interference conditions for all single-hop communication networks of arbitrary large size.

In the following, we use the notations and definitions given in Part I of our multi-part paper [1], almost all of them are standard. Also, channel models and information theoretic concepts such as capacity region are given as usual. Details can be found in [1].

In Section II, we present our new technical lemmas. In Section III, we discuss a part of the ideas for our derivations. In Section IV, we derive our main results for three-user CIC and finally in Section V, we extend the results to the K-user CIC.

[1]Reza K. Farsani was with the department of electrical engineering, Sharif University of Technology. He is by now with the school of cognitive sciences, Institute for Research in Fundamental Sciences (IPM), Tehran, Iran.

# II. New Lemmas

In what follows, we derive some technical lemmas which are repeatedly used throughout the paper. The following results indeed have a central role for our derivations.

***Lemma 1)*** *Let $\mathcal{Y}_1,\mathcal{Y}_2,\mathcal{X}_1,\mathcal{X}_2,\ldots,\mathcal{X}_{\mu_1},\mathcal{X}_{\mu_1+1},\ldots,\mathcal{X}_{\mu_1+\mu_2}$ be arbitrary sets, where $\mu_1,\mu_2\in\mathbb{N}$ are arbitrary natural numbers. Let also $\mathbb{P}(y_1,y_2|x_1,x_2,\ldots,x_{\mu_1},x_{\mu_1+1},\ldots,x_{\mu_1+\mu_2})$ be a given conditional probability distribution defined on the set $\mathcal{Y}_1\times\mathcal{Y}_2\times\mathcal{X}_1\times\mathcal{X}_2\times\ldots\times\mathcal{X}_{\mu_1}\times\mathcal{X}_{\mu_1+1}\times\ldots\times\mathcal{X}_{\mu_1+\mu_2}$. Consider the inequality below:*

$$I(X_1,\ldots,X_{\mu_1};Y_1|X_{\mu_1+1},\ldots,X_{\mu_1+\mu_2})\leq I(X_1,\ldots,X_{\mu_1};Y_2|X_{\mu_1+1},\ldots,X_{\mu_1+\mu_2}) \tag{1}$$

*If the inequality* (1) *holds for all PDFs $P_{X_1\ldots X_{\mu_1}X_{\mu_1+1}\ldots X_{\mu_1+\mu_2}}(x_1,\ldots,x_{\mu_1},x_{\mu_1+1},\ldots,x_{\mu_1+\mu_2})$ with the following factorization:*

$$P_{X_1\ldots X_{\mu_1}X_{\mu_1+1}\ldots X_{\mu_1+\mu_2}}=P_{X_1\ldots X_{\mu_1}}(x_1,\ldots,x_{\mu_1})P_{X_{\mu_1+1}}(x_{\mu_1+1})P_{X_{\mu_1+2}}(x_{\mu_1+2})\ldots P_{X_{\mu_1+\mu_2}}(x_{\mu_1+\mu_2}) \tag{2}$$

*then, we have:*

$$I(X_1,\ldots,X_{\mu_1};Y_1|X_{\mu_1+1},\ldots,X_{\mu_1+\mu_2},D)\leq I(X_1,\ldots,X_{\mu_1};Y_2|X_{\mu_1+1},\ldots,X_{\mu_1+\mu_2},D) \tag{3}$$

*for all joint PDFs $P_{DX_1\ldots X_{\mu_1}X_{\mu_1+1}\ldots X_{\mu_1+\mu_2}}(d,x_1,\ldots,x_{\mu_1},x_{\mu_1+1},\ldots,x_{\mu_1+\mu_2})$ where $D\to X_1,\ldots,X_{\mu_1},X_{\mu_1+1},\ldots,X_{\mu_1+\mu_2}\to Y_1,Y_2$ forms a Markov chain.*

*Proof of Lemma 1)* First we show that (1) implies the following inequality:

$$I(X_1,\ldots,X_{\mu_1};Y_1|X_{\mu_1+1},\ldots,X_{\mu_1+\mu_2},W)\leq I(X_1,\ldots,X_{\mu_1};Y_2|X_{\mu_1+1},\ldots,X_{\mu_1+\mu_2},W) \tag{4}$$

for all PDFs $P_{WX_1\ldots X_{\mu_1}X_{\mu_1+1}\ldots X_{\mu_1+\mu_2}}(w,x_1,\ldots,x_{\mu_1},x_{\mu_1+1},\ldots,x_{\mu_1+\mu_2})$ with:

$$P_{WX_1\ldots X_{\mu_1}X_{\mu_1+1}\ldots X_{\mu_1+\mu_2}}=P_WP_{X_1\ldots X_{\mu_1}|W}P_{X_{\mu_1+1}|W}P_{X_{\mu_1+2}|W}\cdots P_{X_{\mu_1+\mu_2}|W} \tag{5}$$

where $W\to X_1,\ldots,X_{\mu_1},X_{\mu_1+1},\ldots,X_{\mu_1+\mu_2}\to Y_1,Y_2$ forms a Markov chain. To prove this inequality, one can write:

$$\begin{aligned}
&I(X_1,\ldots,X_{\mu_1};Y_1|X_{\mu_1+1},\ldots,X_{\mu_1+\mu_2},W)\\
&\quad=\sum_w P_W(w)I(X_1,\ldots,X_{\mu_1};Y_1|X_{\mu_1+1},\ldots,X_{\mu_1+\mu_2},w)\\
&\quad=\sum_w P_W(w)I(X_1,\ldots,X_{\mu_1};Y_1|X_{\mu_1+1},\ldots,X_{\mu_1+\mu_2})_{\langle P_{X_1\ldots X_{\mu_1}|w}\times P_{X_{\mu_1+1}|w}\times P_{X_{\mu_1+2}|w}\times\ldots\times P_{X_{\mu_1+\mu_2}|w}\rangle}\\
&\quad\overset{(a)}{\leq}\sum_w P_W(w)I(X_1,\ldots,X_{\mu_1};Y_2|X_{\mu_1+1},\ldots,X_{\mu_1+\mu_2})_{\langle P_{X_1\ldots X_{\mu_1}|w}\times P_{X_{\mu_1+1}|w}\times P_{X_{\mu_1+2}|w}\times\ldots\times P_{X_{\mu_1+\mu_2}|w}\rangle}\\
&\quad=\sum_w P_W(w)I(X_1,\ldots,X_{\mu_1};Y_2|X_{\mu_1+1},\ldots,X_{\mu_1+\mu_2},w)\\
&\quad=I(X_1,\ldots,X_{\mu_1};Y_2|X_{\mu_1+1},\ldots,X_{\mu_1+\mu_2},W)
\end{aligned} \tag{6}$$

where the notation $I(A;B|C)_{\langle P(.)\rangle}$ indicates that the mutual information function $I(A;B|C)$ is evaluated by the distribution $P(.)$. Note that for each certain $w$, the function $P_{X_1\dots X_{\mu_1}|w}\times P_{X_{\mu_1+1}|w}\times P_{X_{\mu_1+2}|w}\times\dots\times P_{X_{\mu_1+\mu_2}|w}$ is a probability distribution defined over the set $\mathcal{X}_1\times\dots\times\mathcal{X}_{\mu_1}\times\mathcal{X}_{\mu_1+1}\times\dots\times\mathcal{X}_{\mu_1+\mu_2}$ with the factorization (2). The inequality (a) is due to (1).

Now, having at hand the inequality (4), one can substitute $W\equiv\left(D,X_{\mu_1+1},X_{\mu_1+2},\dots,X_{\mu_1+\mu_2}\right)$ with an arbitrary joint distribution[2] on the set $\mathcal{D}\times\mathcal{X}_{\mu_1+1}\times\dots\times\mathcal{X}_{\mu_1+\mu_2}$ . By this substitution, we derive that (3) holds for all joint PDFs $P_{DX_{\mu_1+1}\dots X_{\mu_1+\mu_2}}P_{X_1\dots X_{\mu_1}|DX_{\mu_1+1}\dots X_{\mu_1+\mu_2}}$. The proof is complete. ■

***Remark 1)*** It is essential to remark that the inequality (3) holds only for those auxiliary random variables "$D$" where $D\to X_1,\dots,X_{\mu_1},X_{\mu_1+1},\dots,X_{\mu_1+\mu_2}\to Y_1,Y_2$ forms a Markov chain. For example, it is wrong to set $D\equiv Y_2$ in (3) and deduce that:

$$I\left(X_1,\dots,X_{\mu_1};Y_1\middle|X_{\mu_1+1},\dots,X_{\mu_1+\mu_2},Y_2\right)\overset{?}{=}0 \tag{7}$$

In general, (1) *does not* imply the equality (7). For more explanation, consider a two-user broadcast channel with input $X$ and outputs $Y_1$ and $Y_2$ with transition probability function $\mathbb{P}(y_1,y_2|x)$. Consider the following condition:

$$I(X;Y_2)\le I(X;Y_1)\qquad \textit{for all joint PDFs}\qquad P_X(x) \tag{8}$$

The condition (8) indeed represents the class of *more-capable* broadcast channels [12]. According to Lemma 1, (8) imply that:

$$I(X;Y_2|D)\le I(X;Y_1|D)\qquad \textit{for all joint PDFs}\qquad P_{DX}(d,x) \tag{9}$$

Let us now set $D\equiv Y_1$ in (9). We obtain that $I(X;Y_2|Y_1)=0$. It is clear that the latter equality implies that $X\to Y_1\to Y_2$ forms a Markov chain, i.e., $Y_2$ is a degraded version of $Y_1$. But we know [12] that the more-capable broadcast channels strictly include the degraded BCs as a subset. In other words, the condition (8) in general does not imply that $I(X;Y_2|Y_1)=0$. The fact is that in (9) it is required that $D\to X\to Y_1,Y_2$ forms a Markov chain. Therefore, the choice $D\equiv Y_1$ is not admissible.

***Corollary 1)*** *Let $\mathcal{L}$ be an arbitrary subset of $\{1,\dots,\mu_1\}$. Denote $\mathbb{X}_{\mathcal{L}}\triangleq\{X_i : i\in\mathcal{L}\}$. If the inequality* (1) *holds for all joint PDFs* (2), *then we have:*

$$I\left(\{X_1,\dots,X_{\mu_1}\}-\mathbb{X}_{\mathcal{L}};Y_1\middle|\mathbb{X}_{\mathcal{L}},X_{\mu_1+1},\dots,X_{\mu_1+\mu_2},D\right)\le I\left(\{X_1,\dots,X_{\mu_1}\}-\mathbb{X}_{\mathcal{L}};Y_2\middle|\mathbb{X}_{\mathcal{L}},X_{\mu_1+1},\dots,X_{\mu_1+\mu_2},D\right) \tag{10}$$

*for all joint PDFs $P_{DX_1\dots X_{\mu_1}X_{\mu_1+1}\dots X_{\mu_1+\mu_2}}\left(d,x_1,\dots,x_{\mu_1},x_{\mu_1+1},\dots,x_{\mu_1+\mu_2}\right)$ where $D\to X_1,\dots,X_{\mu_1},X_{\mu_1+1},\dots,X_{\mu_1+\mu_2}\to Y_1,Y_2$ forms a Markov chain.*

*Proof of Corollary 1)* It is sufficient to replace $D$ with $(D,\mathbb{X}_{\mathcal{L}})$ in (3). ■

Next let us consider a Gaussian transition probability function. Precisely, let the outputs $Y_1$ and $Y_2$ be given as follows:

$$\begin{cases}Y_1\triangleq a_1X_1+a_2X_2+\dots+a_{\mu_1}X_{\mu_1}+a_{\mu_1+1}X_{\mu_1+1}+\dots+a_{\mu_1+\mu_2}X_{\mu_1+\mu_2}+Z_1\\ Y_2\triangleq b_1X_1+b_2X_2+\dots+b_{\mu_1}X_{\mu_1}+b_{\mu_1+1}X_{\mu_1+1}+\dots+b_{\mu_1+\mu_2}X_{\mu_1+\mu_2}+Z_2\end{cases} \tag{11}$$

where $Z_1$ and $Z_2$ are zero-mean unit-variance Gaussian random variables; also, $X_1,X_2,\dots,X_{\mu_1},X_{\mu_1+1},\dots,X_{\mu_1+\mu_2}$ are real-valued power-constrained random variables independent of $(Z_1,Z_2)$ and $a_1,a_2,\dots,a_{\mu_1},a_{\mu_1+1},\dots,a_{\mu_1+\mu_2}$ and $b_1,b_2,\dots,b_{\mu_1},b_{\mu_1+1},\dots,b_{\mu_1+\mu_2}$ are fixed real numbers. Our purpose is to determine sufficient conditions under which for this setup the inequality (3) holds for all joint PDFs $P_{DX_1\dots X_{\mu_1}X_{\mu_1+1}\dots X_{\mu_1+\mu_2}}(d,x_1,\dots,x_{\mu_1},x_{\mu_1+1},\dots,x_{\mu_1+\mu_2})$. The following lemma gives such conditions.

[2] We have this liberty because $P_W(w)$ in (5) is arbitrary.

***Lemma 2)*** *Consider the Gaussian system in* (11). *If the following condition satisfies:*

$$\frac{a_1}{b_1}=\frac{a_2}{b_2}=\cdots=\frac{a_{\mu_1}}{b_{\mu_1}}=\alpha,\qquad |\alpha|\leq 1 \tag{12}$$

*then, the inequality* (3) *holds for all joint PDFs* $P_{DX_1\dots X_{\mu_1}X_{\mu_1+1}\dots X_{\mu_1+\mu_2}}(d,x_1,\dots,x_{\mu_1},x_{\mu_1+1},\dots,x_{\mu_1+\mu_2})$ *where* $D$ *is independent of* $(Z_1,Z_2)$.

*Proof of Lemma 2)* First note that if $D$ is independent of $(Z_1,Z_2)$, then $D\to X_1,\dots,X_{\mu_1},X_{\mu_1+1},\dots,X_{\mu_1+\mu_2}\to Y_1,Y_2$ forms a Markov chain. It is sufficient to prove that (1) holds. Define:

$$\tilde{Y}_1\triangleq \alpha Y_2+(\alpha b_{\mu_1+1}-a_{\mu_1+1})X_{\mu_1+1}+(\alpha b_{\mu_1+2}-a_{\mu_1+2})X_{\mu_1+2}+\cdots+(\alpha b_{\mu_1+\mu_2}-a_{\mu_1+\mu_2})X_{\mu_1+\mu_2}+\sqrt{1-\alpha^2}\tilde{Z}_1 \tag{13}$$

where $\tilde{Z}_1$ is a Gaussian random variable with zero mean and unit variance and impendent of $(Z_1,Z_2)$. Considering (11), it is readily derived that $\tilde{Y}_1$ is statistically equivalent to $Y_1$ in the sense of:

$$\mathbb{P}(\tilde{y}_1|x_1,\dots,x_{\mu_1},x_{\mu_1+1},\dots,x_{\mu_1+\mu_2})\approxeq\mathbb{P}(y_1|x_1,\dots,x_{\mu_1},x_{\mu_1+1},\dots,x_{\mu_1+\mu_2})$$

Therefore, for all input distributions we have:

$$\begin{aligned}I(X_1,\dots,X_{\mu_1};Y_1|X_{\mu_1+1},\dots,X_{\mu_1+\mu_2})&=I(X_1,\dots,X_{\mu_1};\tilde{Y}_1|X_{\mu_1+1},\dots,X_{\mu_1+\mu_2})\\&\leq I(X_1,\dots,X_{\mu_1};\tilde{Y}_1,Y_2|X_{\mu_1+1},\dots,X_{\mu_1+\mu_2})\\&\overset{(a)}{=}I(X_1,\dots,X_{\mu_1};Y_2|X_{\mu_1+1},\dots,X_{\mu_1+\mu_2})+I(X_1,\dots,X_{\mu_1};\tilde{Y}_1|Y_2,X_{\mu_1+1},\dots,X_{\mu_1+\mu_2})\\&=I(X_1,\dots,X_{\mu_1};Y_2|X_{\mu_1+1},\dots,X_{\mu_1+\mu_2})\end{aligned}$$

where (a) holds because, according to (13), $X_1,\dots,X_{\mu_1}\to Y_2,X_{\mu_1+1},\dots,X_{\mu_1+\mu_2}\to\tilde{Y}_1$ forms a Markov chain. The proof is complete. ■

***Remarks 2:***

1. The proof style of Lemma 2 indicates that under the condition (12), given $X_{\mu_1+1},\dots,X_{\mu_1+\mu_2}$, the signal $Y_1$ is a stochastically degraded version of $Y_2$.
2. The relation (12) is a sufficient condition under which (1) holds; however, in general the inequality (1) may not be equivalent to (12). It is also essential to note that the condition (12) is not derived by evaluating (1) for Gaussian input distributions. Only for the case of $\mu_1=1$, the condition (12) can be equivalently derived by evaluating (1) for Gaussian input distributions.

In the next lemma, we also provide a multi-letter extension of lemma 1 which is necessary to identify strong interference regime for multi-user networks.

***Lemma 3)*** *Fix the conditional PDF* $\mathbb{P}(y_1,y_2|x_1,\dots,x_{\mu_1},x_{\mu_1+1},\dots,x_{\mu_1+\mu_2})$. *Assume that the inequality* (1) *holds for all joint PDFs* (2). *For a given arbitrary natural number* $n$, *let* $\mathbb{P}(y_1^n,y_2^n|x_1^n,x_2^n,\dots,x_{\mu_1}^n,x_{\mu_1+1}^n,\dots,x_{\mu_1+\mu_2}^n)$ *be a memoryless* $n$*-tuple extension of* $\mathbb{P}(y_1,y_2|x_1,x_2,\dots,x_{\mu_1},x_{\mu_1+1},\dots,x_{\mu_1+\mu_2})$, *i.e.,*

$$\mathbb{P}(y_1^n,y_2^n|x_1^n,x_2^n,\dots,x_{\mu_1}^n,x_{\mu_1+1}^n,\dots,x_{\mu_1+\mu_2}^n)=\prod_{t=1}^{n}\mathbb{P}(y_{1,t},y_{2,t}|x_{1,t},x_{2,t},\dots,x_{\mu_1,t},x_{\mu_1+1,t},\dots,x_{\mu_1+\mu_2,t}) \tag{14}$$

*Then, the following inequality holds:*

$$I\left(X_1^n,\dots,X_{\mu_1}^n;Y_1^n\middle|X_{\mu_1+1}^n,\dots,X_{\mu_1+\mu_2}^n,D\right)\leq I\left(X_1^n,\dots,X_{\mu_1}^n;Y_2^n\middle|X_{\mu_1+1}^n,\dots,X_{\mu_1+\mu_2}^n,D\right) \tag{15}$$

*for all joint PDFs* $P_{DX_1^n\dots X_{\mu_1}^n X_{\mu_1+1}^n\dots X_{\mu_1+\mu_2}^n}\left(d,x_1^n,\dots,x_{\mu_1}^n,x_{\mu_1+1}^n,\dots,x_{\mu_1+\mu_2}^n\right)$ *where* $D\to X_1^n,\dots,X_{\mu_1}^n,X_{\mu_1+1}^n,\dots,X_{\mu_1+\mu_2}^n\to Y_1^n,Y_2^n$ *forms a Markov chain.*

*Proof of Lemma 3)* First note that, according to Lemma 1, since (1) holds for all joint PDFs (2), the inequality (3) also holds for all joint PDFs $P_{DX_1\dots X_{\mu_1}X_{\mu_1+1}\dots X_{\mu_1+\mu_2}}\left(d,x_1,\dots,x_{\mu_1},x_{\mu_1+1},\dots,x_{\mu_1+\mu_2}\right)$ where $D\to X_1,\dots,X_{\mu_1},X_{\mu_1+1},\dots,X_{\mu_1+\mu_2}\to Y_1,Y_2$ forms a Markov chain. Now consider the two sides of (15). For a given vector $A^n$, denote $A^{n\backslash t}\triangleq\left(A^{t-1},A_{t+1}^n\right)$ where $t=1,..,n$. Define:

$$\bar{\bar{D}}\triangleq\left(X_{\mu_1+1}^{n\backslash t},\dots,X_{\mu_1+\mu_2}^{n\backslash t},Y_2^{t-1},Y_{1,t+1}^n,D\right) \tag{16}$$

We have:

$$\begin{aligned}
&I\left(X_1^n,\dots,X_{\mu_1}^n;Y_2^n\middle|X_{\mu_1+1}^n,\dots,X_{\mu_1+\mu_2}^n,D\right)-I\left(X_1^n,\dots,X_{\mu_1}^n;Y_1^n\middle|X_{\mu_1+1}^n,\dots,X_{\mu_1+\mu_2}^n,D\right)\\
&\quad=\sum_{t=1}^n I\left(X_1^n,\dots,X_{\mu_1}^n;Y_{2,t}\middle|X_{\mu_1+1}^n,\dots,X_{\mu_1+\mu_2}^n,Y_2^{t-1},D\right)-\sum_{t=1}^n I\left(X_1^n,\dots,X_{\mu_1}^n;Y_{1,t}\middle|X_{\mu_1+1}^n,\dots,X_{\mu_1+\mu_2}^n,Y_{1,t+1}^n,D\right)\\
&\quad\overset{(a)}{=}\sum_{t=1}^n I\left(X_{1,t},\dots,X_{\mu_1,t};Y_{2,t}\middle|X_{\mu_1+1}^n,\dots,X_{\mu_1+\mu_2}^n,Y_2^{t-1},D\right)-\sum_{t=1}^n I\left(X_{1,t},\dots,X_{\mu_1,t};Y_{1,t}\middle|X_{\mu_1+1}^n,\dots,X_{\mu_1+\mu_2}^n,Y_{1,t+1}^n,D\right)\\
&\quad\overset{(b)}{=}\sum_{t=1}^n I\left(X_{1,t},\dots,X_{\mu_1,t},Y_{1,t+1}^n;Y_{2,t}\middle|X_{\mu_1+1}^n,\dots,X_{\mu_1+\mu_2}^n,Y_2^{t-1},D\right)-\sum_{t=1}^n I\left(X_{1,t},\dots,X_{\mu_1,t},Y_2^{t-1};Y_{1,t}\middle|X_{\mu_1+1}^n,\dots,X_{\mu_1+\mu_2}^n,Y_{1,t+1}^n,D\right)\\
&\quad=\sum_{t=1}^n I\left(X_{1,t},\dots,X_{\mu_1,t};Y_{2,t}\middle|X_{\mu_1+1}^n,\dots,X_{\mu_1+\mu_2}^n,Y_{1,t+1}^n,Y_2^{t-1},D\right)-\sum_{t=1}^n I\left(X_{1,t},\dots,X_{\mu_1,t};Y_{1,t}\middle|X_{\mu_1+1}^n,\dots,X_{\mu_1+\mu_2}^n,Y_{1,t+1}^n,Y_2^{t-1},D\right)\\
&\qquad\qquad+\sum_{t=1}^n I\left(Y_{1,t+1}^n;Y_{2,t}\middle|X_{\mu_1+1}^n,\dots,X_{\mu_1+\mu_2}^n,Y_2^{t-1},D\right)-\sum_{t=1}^n I\left(Y_2^{t-1};Y_{1,t}\middle|X_{\mu_1+1}^n,\dots,X_{\mu_1+\mu_2}^n,Y_{1,t+1}^n,D\right)\\
&\quad\overset{(c)}{=}\sum_{t=1}^n I\left(X_{1,t},\dots,X_{\mu_1,t};Y_{2,t}\middle|X_{\mu_1+1}^n,\dots,X_{\mu_1+\mu_2}^n,Y_{1,t+1}^n,Y_2^{t-1},D\right)-\sum_{t=1}^n I\left(X_{1,t},\dots,X_{\mu_1,t};Y_{1,t}\middle|X_{\mu_1+1}^n,\dots,X_{\mu_1+\mu_2}^n,Y_{1,t+1}^n,Y_2^{t-1},D\right)\\
&\quad=\sum_{t=1}^n\left(I\left(X_{1,t},\dots,X_{\mu_1,t};Y_{2,t}\middle|X_{\mu_1+1,t},\dots,X_{\mu_1+\mu_2,t},\bar{\bar{D}}\right)-I\left(X_{1,t},\dots,X_{\mu_1,t};Y_{1,t}\middle|X_{\mu_1+1,t},\dots,X_{\mu_1+\mu_2,t},\bar{\bar{D}}\right)\right)\\
&\quad\overset{(d)}{\geq}0
\end{aligned} \tag{17}$$

where equality (a) and (b) hold because the memorylessness property (14) implies the following Markov relations:

$$X_1^{n\backslash t},\dots,X_{\mu_1}^{n\backslash t},X_{\mu_1+1}^{n\backslash t},\dots,X_{\mu_1+\mu_2}^{n\backslash t},Y_2^{t-1},Y_{1,t+1}^n,D\to X_{1,t},\dots,X_{\mu_1,t},X_{\mu_1+1,t},\dots,X_{\mu_1+\mu_2,t}\to Y_{1,t},Y_{2,t},\qquad t=1,\dots,n \tag{18}$$

Also, equality (c) is due to Csiszar-Korner identity according which the $3^{rd}$ and the $4^{th}$ expressions in the left hand side of (c) are equal; finally, (d) is due to the inequality (3) in which $D$ is replaced by $\bar{\bar{D}}$. The proof is complete. ■

Let us consider the special case of $\mu_1=\mu_2=2$. In [11, Appendix], it is shown that if the following condition holds:

$$I(X_1;Y_1|X_2) \leq I(X_1;Y_2|X_2) \quad \text{for all joint PDFs} \quad P_{X_1}P_{X_2} \tag{19}$$

then, we have:

$$I(X_1^n;Y_1^n|X_2^n) \leq I(X_1^n;Y_2^n|X_2^n) \quad \text{for all joint PDFs} \quad P_{X_1^n}P_{X_2^n} \tag{20}$$

The proof of [11] is based on induction which requires establishing some sophisticated Markov chains (see [11, Appendix]). Moreover, the authors of [11] are able to derive (20) only for product distributions $P_{X_1^n} \times P_{X_2^n}$. Our proof in Lemma 3 is considerably simple since, instead of sophisticated induction-based arguments, it is derived by a direct application of the Csiszar-Korner identity. Also, by using the consequence of Lemma 1, we are able to prove (20) for all arbitrary joint PDFs $P_{X_1^nX_2^n}$. As we will see throughout the paper, such extension is critical while deriving strong interference regime for large multi-user networks.

We also derive a variation of Lemma 1 which is useful to identify networks with a sequence of less noisy receivers (these networks are studied in details in Part IV of our multi-part paper [4]). This result is given in the next lemma.

***Lemma 4)*** *Let $\mathcal{Y}_1,\mathcal{Y}_2,\mathcal{X}_1,\mathcal{X}_2,\ldots,\mathcal{X}_{\mu_1},\mathcal{X}_{\mu_1+1},\ldots,\mathcal{X}_{\mu_1+\mu_2}$ be arbitrary sets, where $\mu_1,\mu_2 \in \mathbb{N}$ are arbitrary natural numbers. Let also $\mathbb{P}(y_1,y_2|x_1,x_2,\ldots,x_{\mu_1},x_{\mu_1+1},\ldots,x_{\mu_1+\mu_2})$ be a given conditional probability distribution defined on the set $\mathcal{Y}_1 \times \mathcal{Y}_2 \times \mathcal{X}_1 \times \mathcal{X}_2 \times \ldots \times \mathcal{X}_{\mu_1} \times \mathcal{X}_{\mu_1+1} \times \ldots \times \mathcal{X}_{\mu_1+\mu_2}$. Consider the inequality below:*

$$I(U;Y_1|X_{\mu_1+1},\ldots,X_{\mu_1+\mu_2}) \leq I(U;Y_2|X_{\mu_1+1},\ldots,X_{\mu_1+\mu_2}) \tag{21}$$

*If the inequality* (21) *holds for all PDFs $P_{UX_1\ldots X_{\mu_1}X_{\mu_1+1}\ldots X_{\mu_1+\mu_2}}(x_1,\ldots,x_{\mu_1},x_{\mu_1+1},\ldots,x_{\mu_1+\mu_2})$ with the following factorization:*

$$P_{UX_1\ldots X_{\mu_1}X_{\mu_1+1}\ldots X_{\mu_1+\mu_2}} = P_{UX_1\ldots X_{\mu_1}}(u,x_1,\ldots,x_{\mu_1})P_{X_{\mu_1+1}}(x_{\mu_1+1})P_{X_{\mu_1+2}}(x_{\mu_1+2})\ldots P_{X_{\mu_1+\mu_2}}(x_{\mu_1+\mu_2}) \tag{22}$$

*then, we have:*

$$I(U;Y_1|X_{\mu_1+1},\ldots,X_{\mu_1+\mu_2},D) \leq I(U;Y_2|X_{\mu_1+1},\ldots,X_{\mu_1+\mu_2},D) \tag{23}$$

*for all joint PDFs $P_{DUX_1\ldots X_{\mu_1}X_{\mu_1+1}\ldots X_{\mu_1+\mu_2}}(d,u,x_1,\ldots,x_{\mu_1},x_{\mu_1+1},\ldots,x_{\mu_1+\mu_2})$ where $D,U \to X_1,\ldots,X_{\mu_1},X_{\mu_1+1},\ldots,X_{\mu_1+\mu_2} \to Y_1,Y_2$ forms a Markov chain.*

*Proof of Lemma 4)* The proof is rather similar to Lemma 1. First, note that (21) implies the following inequality:

$$I(U;Y_1|X_{\mu_1+1},\ldots,X_{\mu_1+\mu_2},W) \leq I(U;Y_2|X_{\mu_1+1},\ldots,X_{\mu_1+\mu_2},W) \tag{24}$$

for all PDFs $P_{WUX_1\ldots X_{\mu_1}X_{\mu_1+1}\ldots X_{\mu_1+\mu_2}}(w,u,x_1,\ldots,x_{\mu_1},x_{\mu_1+1},\ldots,x_{\mu_1+\mu_2})$ with:

$$P_{WUX_1\ldots X_{\mu_1}X_{\mu_1+1}\ldots X_{\mu_1+\mu_2}} = P_W P_{UX_1\ldots X_{\mu_1}|W} P_{X_{\mu_1+1}|W} P_{X_{\mu_1+2}|W} \ldots P_{X_{\mu_1+\mu_2}|W} \tag{25}$$

where $W,U \to X_1,\ldots,X_{\mu_1},X_{\mu_1+1},\ldots,X_{\mu_1+\mu_2} \to Y_1,Y_2$ forms a Markov chain. This can be proved by following the same lines as (6). Now, having at hand the inequality (25), one can substitute $W \equiv (X_{\mu_1+1},X_{\mu_1+2},\ldots,X_{\mu_1+\mu_2},D)$ with an arbitrary joint distribution on the set $\mathcal{X}_{\mu_1+1} \times \ldots \times \mathcal{X}_{\mu_1+\mu_2} \times \mathcal{D}$ . By this substitution, we obtain that the inequality (23) holds for all joint PDFs $P_{X_{\mu_1+1}\ldots X_{\mu_1+\mu_2}D}P_{UX_1\ldots X_{\mu_1}|X_{\mu_1+1}\ldots X_{\mu_1+\mu_2}D}$. The proof is complete. ■

Finally, similar to Lemma 2, one can derive sufficient conditions under which for the Gaussian system (11) the inequality (23) holds for all joint PDFs $P_{DUX_1\ldots X_{\mu_1}X_{\mu_1+1}\ldots X_{\mu_1+\mu_2}}(d,u,x_1,\ldots,x_{\mu_1},x_{\mu_1+1},\ldots,x_{\mu_1+\mu_2})$. We present such conditions in the following lemma.

***Lemma 5)*** *Consider the Gaussian system in* (11). *If* (12) *holds, then the inequality* (23) *is satisfied for all joint PDFs* $P_{DUX_1\ldots X_{\mu_1}X_{\mu_1+1}\ldots X_{\mu_1+\mu_2}}(d,u,x_1,\ldots,x_{\mu_1},x_{\mu_1+1},\ldots,x_{\mu_1+\mu_2})$ *where* $(D,U)$ *is independent of* $(Z_1,Z_2)$.

*Proof of Lemma 5)* The proof is indeed similar to Lemma 2. In essence, if the condition (12) holds, given $X_{\mu_1+1},\ldots,X_{\mu_1+\mu_2}$, the signal $Y_1$ is a stochastically degraded version of $Y_2$. This fact was previously indicated in Remark 2. Therefore, (23) is always satisfied for all joint PDFs $P_{DUX_1\ldots X_{\mu_1}X_{\mu_1+1}\ldots X_{\mu_1+\mu_2}}(d,u,x_1,\ldots,x_{\mu_1},x_{\mu_1+1},\ldots,x_{\mu_1+\mu_2})$. ■

***Remark 3)*** Lemma 5 provides only *sufficient* conditions for the Gaussian system (11) to satisfy the inequality (23) for all input distributions.

By these preliminaries, we are ready to develop our results in the subsequent sections.

# III. AN INTERESTING RELATED SCENARIO

Before establishing our main results for the multi-user CICs, let us first discuss an interesting related scenario. Consider a multi-user broadcast channel as shown in Fig. 1.

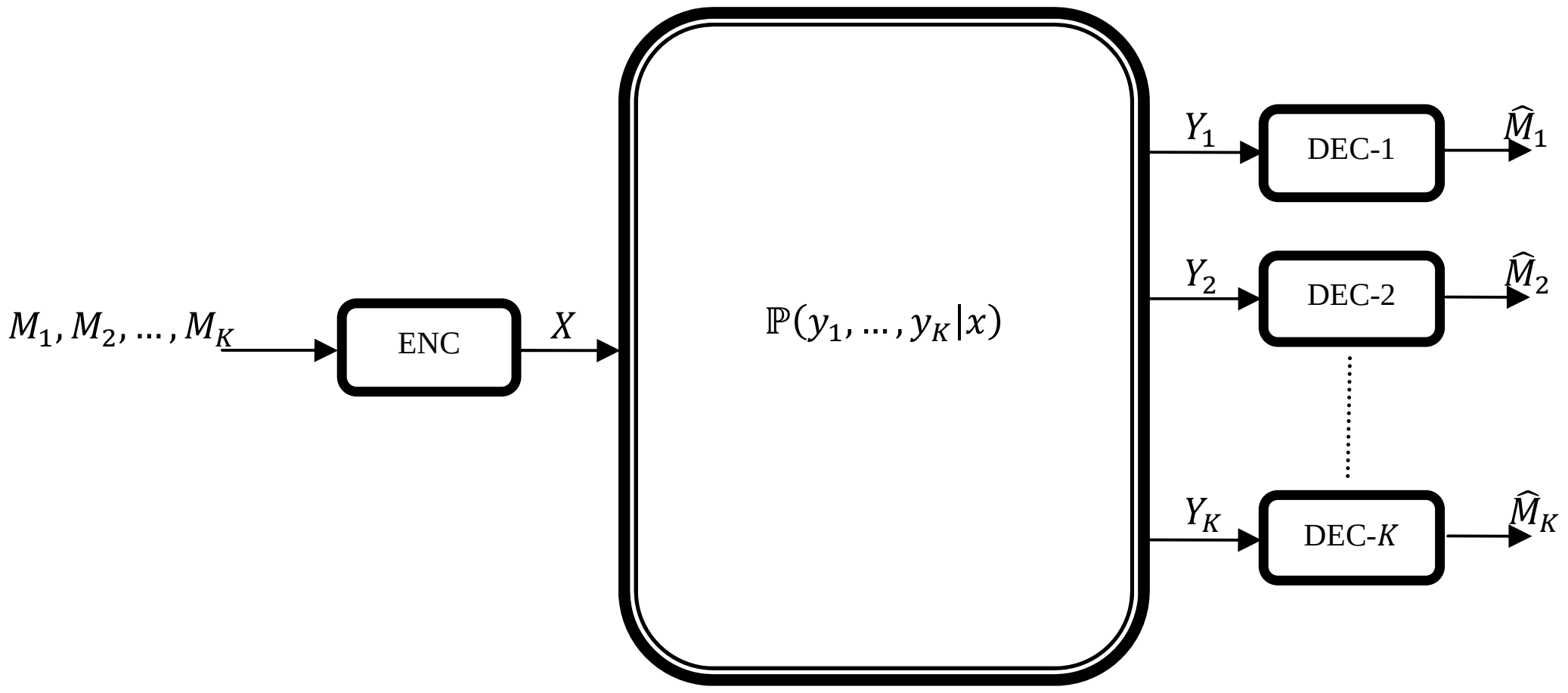

Figure 1. The K-user broadcast channel.

Finding the capacity region for the broadcast networks is one the most difficult problems in network information theory, specifically in the case of more than two receivers. One of important classes for which the capacity region is known in the two-user case is the more-capable channels [12]. A two-user broadcast channel with receivers $Y_1$ and $Y_2$ is said to be more-capable if the condition (8) holds. For this channel, the superposition coding scheme achieves the capacity region, as given in Part I of our multi-part paper [1, Proposition III.3]. As a natural generalization, one may consider the multi-user broadcast channel with sequentially more-capable receivers. Precisely, inspired by the two-user more-capable channel in (8), one may define a multi-user broadcast channel with a sequence of more-capable receivers to be a channel for which the condition below holds:

$$I(X;Y_K) < I(X;Y_{K-1}) < I(X;Y_{K-2}) < \cdots < I(X;Y_1) \quad \text{for all joint PDFs} \quad P_X \tag{26}$$

Unfortunately, the capacity result for the two-user more-capable broadcast channel does not seem to be straightforwardly generalized to the multi-user case in (26). Nonetheless, we could find an insightful result in this regard. Clearly, we find the sum-rate capacity for this network. Let us describe the procedure of the derivation in details. We claim that for the more-capable broadcast channel in (26), the sum-rate capacity is given below:

$$\mathcal{C}_{sum}^{more-capable} = \max_{P_X} I(X;Y_1) \tag{27}$$

The proof of achievability is trivial: the transmitter sends the message $M_1$ for the first user (stronger user) at its capacity rate and withdraws transmission of the other messages. In fact, for our purposes the proof of the converse part is important. Consider a length-$n$ code with vanishing average error probability for the network. First note that, according to Lemma 3, the condition (26) implies that:

$$I(X^n;Y_K^n|D) < I(X^n;Y_{K-1}^n|D) < I(X^n;Y_{K-2}^n|D) < \cdots < I(X^n;Y_1^n|D) \quad \text{for all joint PDFs} \quad P_{DX^n} \tag{28}$$

Now using the Fano's inequality we can write:

$$\begin{aligned}
n\sum_{l=1}^{K} R_l &\leq I(M_K;Y_K^n) + I(M_{K-1};Y_{K-1}^n) + I(M_{K-2};Y_{K-2}^n) + \cdots + I(M_1;Y_1^n) + nK\epsilon_n \\
&\leq I(M_K;Y_K^n, M_{K-1}, M_{K-2}, \dots, M_1) + I(M_{K-1};Y_{K-1}^n, M_{K-2}, \dots, M_1) + I(M_{K-2};Y_{K-2}^n, M_{K-3}, \dots, M_1) + \cdots + I(M_1;Y_1^n) + nK\epsilon_n \\
&= I(M_K;Y_K^n|M_{K-1}, M_{K-2}, \dots, M_1) + I(M_{K-1};Y_{K-1}^n|M_{K-2}, \dots, M_1) + I(M_{K-2};Y_{K-2}^n|M_{K-3}, \dots, M_1) + \cdots + I(M_1;Y_1^n) + nK\epsilon_n \\
&\overset{(a)}{=} I(X^n;Y_K^n|M_{K-1}, M_{K-2}, \dots, M_1) + I(M_{K-1};Y_{K-1}^n|M_{K-2}, \dots, M_1) + I(M_{K-2};Y_{K-2}^n|M_{K-3}, \dots, M_1) + \cdots + I(M_1;Y_1^n) + nK\epsilon_n \\
&\overset{(b)}{\leq} I(X^n;Y_{K-1}^n|M_{K-1}, M_{K-2}, \dots, M_1) + I(M_{K-1};Y_{K-1}^n|M_{K-2}, \dots, M_1) + I(M_{K-2};Y_{K-2}^n|M_{K-3}, \dots, M_1) + \cdots + I(M_1;Y_1^n) + nK\epsilon_n \\
&= I(X^n, M_{K-1};Y_{K-1}^n|M_{K-2}, \dots, M_1) + I(M_{K-2};Y_{K-2}^n|M_{K-3}, \dots, M_1) + \cdots + I(M_1;Y_1^n) + nK\epsilon_n \\
&\overset{(c)}{=} I(X^n;Y_{K-1}^n|M_{K-2}, \dots, M_1) + I(M_{K-2};Y_{K-2}^n|M_{K-3}, \dots, M_1) + \cdots + I(M_1;Y_1^n) + nK\epsilon_n \\
&\overset{(d)}{\leq} I(X^n;Y_{K-2}^n|M_{K-2}, \dots, M_1) + I(M_{K-2};Y_{K-2}^n|M_{K-3}, \dots, M_1) + \cdots + I(M_1;Y_1^n) + nK\epsilon_n \\
&= I(X^n;Y_{K-2}^n|M_{K-3}, \dots, M_1) + \cdots + I(M_1;Y_1^n) + nK\epsilon_n \leq \cdots \leq I(X^n;Y_1^n) + nK\epsilon_n \leq nI(X;Y_1) + nK\epsilon_n
\end{aligned} \tag{29}$$

where $\epsilon_n \to 0$ as $n \to 0$, the equality (a) holds because $X^n$ is given by a deterministic function of $(M_1, M_2, \dots, M_K)$, inequality (b) is due to (28) in which $D$ is replaced by $(M_{K-1}, M_{K-2}, \dots, M_1)$, equality (c) holds because, given the input sequence $X^n$, the output sequence $Y_{K-1}^n$ is independent of messages, inequality (d) is due to (28) and *etc*.

Let us review the philosophy behind the derivations. The first inequality in (29) is a direct consequence of Fano's inequality. The second one is derived by providing (virtual) side information to the receivers. In fact, the messages are sequentially given as side information to the non-respective receivers in a degraded order: $(M_{K-1}, M_{K-2}, \dots, M_1)$ to $Y_K$, $(M_{K-2}, \dots, M_1)$ to $Y_{K-1}$, *etc*. Then, the resulting mutual information functions are successively manipulated (combined) using the more-capable condition in (26) and its extension in Lemma 3, until to reach a single mutual information function. The last mutual information function has a desirable property: it is composed of the input signal and one of the outputs (the stronger one in (26)). This is one of the ideas for our derivations in the following section.

# IV. THE THREE-USER CIC

Consider the three-user CIC shown in Fig. 2.

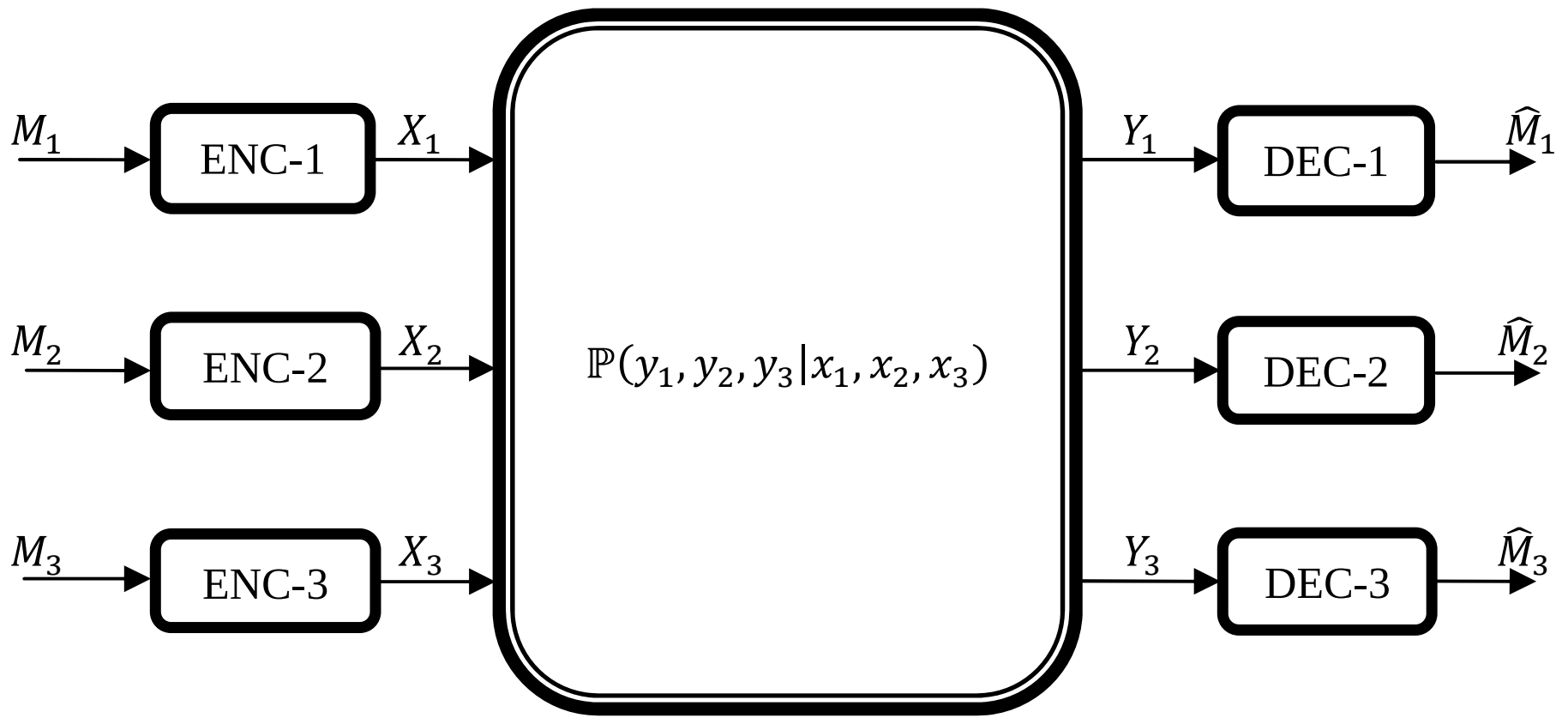

Figure 2. The three-user Classical Interference Channel (CIC).

We intend to derive a strong interference regime for this network. We remark that the following theory which is presented for the three-user CIC can be developed for other interference networks with any arbitrary topology, as given in Part III of our multi-part paper [3, Sec. V.B.3].

First note that according to our definition (see [3, Definition 4]) in the strong interference regime each receiver decodes all messages. The resulting achievable rate region by this scheme is given below:

$$\bigcup_{P_Q P_{X_1|Q} P_{X_2|Q} P_{X_3|Q}} \left\{ \begin{array}{l} (R_1, R_2, R_3): \\ R_1 \leq \min\{I(X_1; Y_1|X_2, X_3, Q), I(X_1; Y_2|X_2, X_3, Q), I(X_1; Y_3|X_2, X_3, Q)\} \\ R_2 \leq \min\{I(X_2; Y_1|X_1, X_3, Q), I(X_2; Y_2|X_1, X_3, Q), I(X_2; Y_3|X_1, X_3, Q)\} \\ R_3 \leq \min\{I(X_3; Y_1|X_1, X_2, Q), I(X_3; Y_2|X_1, X_2, Q), I(X_3; Y_3|X_1, X_2, Q)\} \\ \\ R_1 + R_2 \leq \min\{I(X_1, X_2; Y_1|X_3, Q), I(X_1, X_2; Y_2|X_3, Q), I(X_1, X_2; Y_3|X_3, Q)\} \\ R_2 + R_3 \leq \min\{I(X_2, X_3; Y_1|X_1, Q), I(X_2, X_3; Y_2|X_1, Q), I(X_2, X_3; Y_3|X_1, Q)\} \\ R_1 + R_3 \leq \min\{I(X_1, X_3; Y_1|X_2, Q), I(X_1, X_3; Y_2|X_2, Q), I(X_1, X_3; Y_3|X_2, Q)\} \\ \\ R_1 + R_2 + R_3 \leq \min\{I(X_1, X_2, X_3; Y_1|Q), I(X_1, X_2, X_3; Y_2|Q), I(X_1, X_2, X_3; Y_3|Q)\} \end{array} \right\} \tag{30}$$

We need to derive conditions under which this rate region is optimal. Consider a length-$n$ block code for the network with vanishing error probability.

*Claim:* if the network transition probability function satisfies the following conditions:

$$\begin{cases} I(X_2; Y_2|X_1, X_3) \leq I(X_2; Y_3|X_1, X_3) & \text{for all joint PDFs} \quad P_{X_1}P_{X_2}P_{X_3} \\ I(X_2, X_3; Y_3|X_1) \leq I(X_2, X_3; Y_1|X_1) & \text{for all joint PDFs} \quad P_{X_1}P_{X_2X_3} \end{cases} \tag{31}$$

then, we have:

$$n(R_1 + R_2 + R_3) \leq I(X_1^n, X_2^n, X_3^n; Y_1^n) + n\epsilon_n \leq \sum_{t=1}^{n} I\left(X_{1,t}, X_{2,t}, X_{3,t}; Y_{1,t}\right) + n\epsilon_n \tag{32}$$

where $\epsilon_n \to 0$ as $n \to 0$.

*Proof of Claim:* Based on the Fano's inequality one can write:

$$n(R_1+R_2+R_3) \leq I(M_2;Y_2^n)+I(M_3;Y_3^n)+I(M_1;Y_1^n)+n\epsilon_n$$
$$\leq I(M_2;Y_2^n|M_1,M_3)+I(M_3;Y_3^n|M_1)+I(M_1;Y_1^n)+n\epsilon_n$$
$$\overset{(a)}{=} I(X_2^n;Y_2^n|X_1^n,X_3^n,M_1,M_3)+I(X_3^n,M_3;Y_3^n|X_1^n,M_1)+I(X_1^n,M_1;Y_1^n)+n\epsilon_n$$
$$\overset{(b)}{\leq} I(X_2^n;Y_3^n|X_1^n,X_3^n,M_1,M_3)+I(X_3^n,M_3;Y_3^n|X_1^n,M_1)+I(X_1^n,M_1;Y_1^n)+n\epsilon_n$$
$$= I(X_2^n,X_3^n;Y_3^n|X_1^n,M_1)+I(X_1^n,M_1;Y_1^n)+n\epsilon_n$$
$$\overset{(c)}{\leq} I(X_2^n,X_3^n;Y_1^n|X_1^n,M_1)+I(X_1^n,M_1;Y_1^n)+n\epsilon_n$$
$$= I(X_1^n,X_2^n,X_3^n;Y_1^n)+n\epsilon_n \leq \sum_{t=1}^{n} I\big(X_{1,t},X_{2,t},X_{3,t};Y_{1,t}\big)+n\epsilon_n \tag{33}$$

where equality (a) holds because the input sequence $X_i^n$ is given by a deterministic function of the message $M_i, i=1,2,3$, inequality (b) is due to the first condition in (31) and its $n$-tuple extension in Lemma 3, and inequality (c) is due to the second condition in (31) and its $n$-tuple extension in Lemma 3.

Therefore, under the conditions (31), we derived one of the desired constraints on the sum-rate capacity in (30). Indeed, by the conditions (31), one can achieve further results. Clearly, these conditions imply that decoding all messages at the first receiver is optimal. Let us prove this conclusion. Consider the constraints on the partial sum rates. First note that, according to Corollary 1, the second condition of (31) imply that:

$$\begin{cases} I(X_2;Y_3|X_1,X_3) \leq I(X_2;Y_1|X_1,X_3) \\ I(X_3;Y_3|X_1,X_2) \leq I(X_3;Y_1|X_1,X_2) \end{cases} \tag{34}$$

Comparing the first condition in (31) and the first condition of (34), we also obtain:

$$I(X_2;Y_2|X_1,X_3) \leq I(X_2;Y_1|X_1,X_3) \tag{35}$$

Now, we have:

$$n(R_1+R_2) \leq I(M_2;Y_2^n)+I(M_1;Y_1^n)+n\epsilon_n$$
$$\leq I(M_2;Y_2^n|M_1,M_3)+I(M_1;Y_1^n|M_3)+n\epsilon_n$$
$$= I(X_2^n;Y_2^n|X_1^n,X_3^n,M_1,M_3)+I(X_1^n,M_1;Y_1^n|X_3^n,M_3)+n\epsilon_n$$
$$\overset{(a)}{\leq} I(X_2^n;Y_1^n|X_1^n,X_3^n,M_1,M_3)+I(X_1^n,M_1;Y_1^n|X_3^n,M_3)+n\epsilon_n$$
$$= I(X_1^n,X_2^n;Y_1^n|X_3^n,M_3)+n\epsilon_n \leq \sum_{t=1}^{n} I\big(X_{1,t},X_{2,t};Y_{1,t}\big|X_{3,t}\big)+n\epsilon_n \tag{36}$$

where inequality (a) is due to (35) and its $n$-tuple extension in Lemma 3. Also, by following the same lines as (33), one can derive:

$$R_2+R_3 \leq I(X_2^n,X_3^n;Y_1^n|X_1^n,M_1)+n\epsilon_n \leq \sum_{t=1}^{n} I\big(X_{2,t},X_{3,t};Y_{1,t}\big|X_{1,t}\big)+n\epsilon_n \tag{37}$$

Note that (37) actually is the inequality (c) of (33) in which the term $I(X_1^n,M_1;Y_1^n)$ from the right side and the corresponding rate $R_1$ from the left side are removed. Lastly, we have:

$$n(R_1+R_3) \leq I(M_3;Y_3^n)+I(M_1;Y_1^n)+n\epsilon_n$$
$$\leq I(X_3^n;Y_3^n|X_1^n,X_2^n,M_1,M_2)+I(X_1^n,M_1;Y_1^n|X_2^n,M_2)+n\epsilon_n$$

$$\overset{(a)}{\leq} I(X_3^n;Y_1^n|X_1^n,X_2^n,M_1,M_2)+I(X_1^n,M_1;Y_1^n|X_2^n,M_2)+n\epsilon_n$$
$$=I(X_1^n,X_3^n;Y_1^n|X_2^n,M_2)+n\epsilon_n\leq\sum_{t=1}^n I\big(X_{1,t},X_{3,t};Y_{1,t}\big|X_{2,t}\big)+n\epsilon_n$$

(38)

where inequality (a) is due to the second condition in (34). Finally, the desired constraints on the individual rates can be easily derived using the second condition of (34) and the condition (35). Thus, if (31) holds, then it is optimal to decode all messages at the first receiver. It is clear that we can follow the same procedure for the other receivers to derive conditions under which the strong interference criterion, i.e., the optimality of decoding all messages, is satisfied. For example, one can verify that if the following conditions hold:

$$\begin{cases} I(X_3;Y_3|X_1,X_2)\leq I(X_3;Y_1|X_1,X_2) & \text{for all joint PDFs} \quad P_{X_1}P_{X_2}P_{X_3} \\ I(X_1,X_3;Y_1|X_2)\leq I(X_1,X_3;Y_2|X_2) & \text{for all joint PDFs} \quad P_{X_1X_3}P_{X_2} \end{cases}$$

(39)

the second receiver, and if the following hold:

$$\begin{cases} I(X_1;Y_1|X_2,X_3)\leq I(X_1;Y_2|X_2,X_3) & \text{for all joint PDFs} \quad P_{X_1}P_{X_2}P_{X_3} \\ I(X_1,X_2;Y_2|X_3)\leq I(X_1,X_2;Y_3|X_3) & \text{for all joint PDFs} \quad P_{X_1X_2}P_{X_3} \end{cases}$$

(40)

the third receiver experience strong interference. Therefore, the collection of the conditions (31), (39) and (40) constitutes a strong interference regime for the three-user CIC in Fig. 2. A remarkable point is that the necessary conditions for deriving the desired constraints on the sum-rate such as (33) are indeed sufficient to prove the optimality of decoding all messages at the receivers. In other words, once we derived the desired constraints on the sum-rate capacity using certain conditions, no additional condition is required to be introduced to prove the desired constraints on the partial sum-rates.

Let us concentrate on this collection. According to Corollary 1, the second condition of (31) implies the first condition of (39), the second condition of (39) implies the first condition in (40) and the second condition of (40) implies the first condition of (31). Therefore, a strong interference regime for the three-user is given as follows:

**A strong interference regime for the 3-user interference channel**

$$I(X_2,X_3;Y_3|X_1)\leq I(X_2,X_3;Y_1|X_1) \quad \text{for all joint PDFs} \quad P_{X_1}P_{X_2X_3}$$
$$I(X_1,X_3;Y_1|X_2)\leq I(X_1,X_3;Y_2|X_2) \quad \text{for all joint PDFs} \quad P_{X_1X_3}P_{X_2}$$
$$I(X_1,X_2;Y_2|X_3)\leq I(X_1,X_2;Y_3|X_3) \quad \text{for all joint PDFs} \quad P_{X_1X_2}P_{X_3}$$

(41)

The conditions (41) to some extent represents a fact regarding the CICs that is the signal of each receiver is impaired by the joint effect of interference from all non-corresponding transmitters rather by each transmitter's signal separately. Note that the terms in the right side of the inequalities in (41) indeed measure the amount of interference experienced by the receivers.

It should be noted that using the conditions (41) we are able to derive all the constraints in the rate region (30); nevertheless, some of these constraints are actually redundant. In fact, if the conditions (41) hold, the rate region (30) is simplified below:

$$\bigcup_{P_Q P_{X_1|Q} P_{X_2|Q} P_{X_3|Q}} \left\{ \begin{array}{l} (R_1, R_2, R_3): \\ R_1 \leq I(X_1; Y_1 | X_2, X_3, Q) \\ R_2 \leq I(X_2; Y_2 | X_1, X_3, Q) \\ R_3 \leq I(X_3; Y_3 | X_1, X_2, Q) \\ \\ R_1 + R_2 \leq \min\{I(X_1, X_2; Y_1 | X_3, Q), I(X_1, X_2; Y_2 | X_3, Q)\} \\ R_2 + R_3 \leq \min\{I(X_2, X_3; Y_2 | X_1, Q), I(X_2, X_3; Y_3 | X_1, Q)\} \\ R_1 + R_3 \leq \min\{I(X_1, X_3; Y_1 | X_2, Q), I(X_1, X_3; Y_3 | X_2, Q)\} \\ \\ R_1 + R_2 + R_3 \leq \min \left\{ \begin{array}{l} I(X_1, X_2, X_3; Y_1 | Q), \\ I(X_1, X_2, X_3; Y_2 | Q), \\ I(X_1, X_2, X_3; Y_3 | Q) \end{array} \right\} \end{array} \right\}$$

(42)

The other constraints of (30) are relaxed by the conditions in (41). Let us now consider the three-user Gaussian CIC as formulated in the following standard form:

$$\begin{bmatrix} Y_1 \\ Y_2 \\ Y_3 \end{bmatrix} = \begin{bmatrix} 1 & a_{12} & a_{13} \\ a_{21} & 1 & a_{23} \\ a_{31} & a_{32} & 1 \end{bmatrix} \begin{bmatrix} X_1 \\ X_2 \\ X_3 \end{bmatrix} + \begin{bmatrix} Z_1 \\ Z_2 \\ Z_3 \end{bmatrix}$$

(43)

where $Z_1, Z_2, Z_3$ are zero-mean unit-variance Gaussian noises and $\mathbb{E}[X_i^2] \leq P_i,\ i = 1,2,3$. Using Lemma 2, one can derive explicit constraints on the network gains under which the strong interference regime (41) holds, as given below:

$$\left\{ \begin{array}{lll} |a_{13}| \geq 1, & |a_{21}| \geq 1, & |a_{32}| \geq 1 \\ & , & \\ a_{12} = a_{13} a_{32}, & a_{31} = a_{21} a_{32}, & a_{23} = a_{21} a_{13} \end{array} \right.$$

(44)

Let examine the conditions (44). Among six parameters in the network gain matrix (43), the parameters $a_{13}, a_{21}$ and $a_{32}$, which no pair of them lies in either a same row or a same column, are given by arbitrary real numbers greater than one and the other parameters are given in terms of these parameters by specific relations.

Here, we return to the calculations (33) where we derived the constraint (32) on the sum-rate using the conditions (31). If we review these calculations, we observe that by exchanging the order of manipulating mutual information functions, one can derive conditions other than those in (31) under which decoding of all messages at the first receiver is optimal. Specifically, consider the following conditions:

$$\left\{ \begin{array}{lll} I(X_3; Y_3 | X_1, X_2) \leq I(X_3; Y_2 | X_1, X_2) & \text{for all joint PDFs} & P_{X_1} P_{X_2} P_{X_3} \\ \\ I(X_2, X_3; Y_2 | X_1) \leq I(X_2, X_3; Y_1 | X_1) & \text{for all joint PDFs} & P_{X_1} P_{X_2 X_3} \end{array} \right.$$

(45)

The conditions (45) are obtained by exchanging the indices “2” and “3” in (31). One can readily verify the conditions (45) also imply that decoding all messages at the first receiver is optimal. The derivation is similar to (33)-(38) except that the indices “2” and “3” are exchanged everywhere. Therefore, the collection of (45), (39) and (40) is also a strong interference regime for the three-user CIC. This second collection of strong interference conditions can be represented in the form of below:

$$\left\{ \begin{array}{lll} I(X_3; Y_3 | X_1, X_2) \leq I(X_3; Y_2 | X_1, X_2) & \text{for all joint PDFs} & P_{X_1} P_{X_2} P_{X_3} \\ I(X_2, X_3; Y_2 | X_1) \leq I(X_2, X_3; Y_1 | X_1) & \text{for all joint PDFs} & P_{X_1} P_{X_2 X_3} \\ I(X_1, X_3; Y_1 | X_2) \leq I(X_1, X_3; Y_2 | X_2) & \text{for all joint PDFs} & P_{X_1 X_3} P_{X_2} \\ I(X_1, X_2; Y_2 | X_3) \leq I(X_1, X_2; Y_3 | X_3) & \text{for all joint PDFs} & P_{X_1 X_2} P_{X_3} \end{array} \right.$$

(46)

For the Gaussian network in (43) Lemma 2 implies that if the following constraints hold:

$$\begin{cases} 1 \leq |a_{23}| \\ ,\\ a_{21} = \dfrac{1}{a_{12}} = \dfrac{a_{23}}{a_{13}} = \alpha, \qquad |\alpha| = 1 \\ ,\\ \dfrac{a_{21}}{a_{31}} = \dfrac{1}{a_{32}} = \beta, \qquad |\beta| \leq 1 \end{cases} \tag{47}$$

then, the conditions (46) are satisfied.

In fact, by following the same procedure, one can derive $2^3 = 8$ different strong interference regimes for the three-user CIC. Among these regimes, $(3-1)! = 2$ ones are more significant: the regime in (41) and the regime that is derived by exchanging 1 by 3 in (41). The other regimes, for example that in (46), lead to rather trivial situations, specifically for the Gaussian networks.

# V. THE K-USER CIC

Now, let us examine the CIC with arbitrary number of users as shown in Fig. 3.

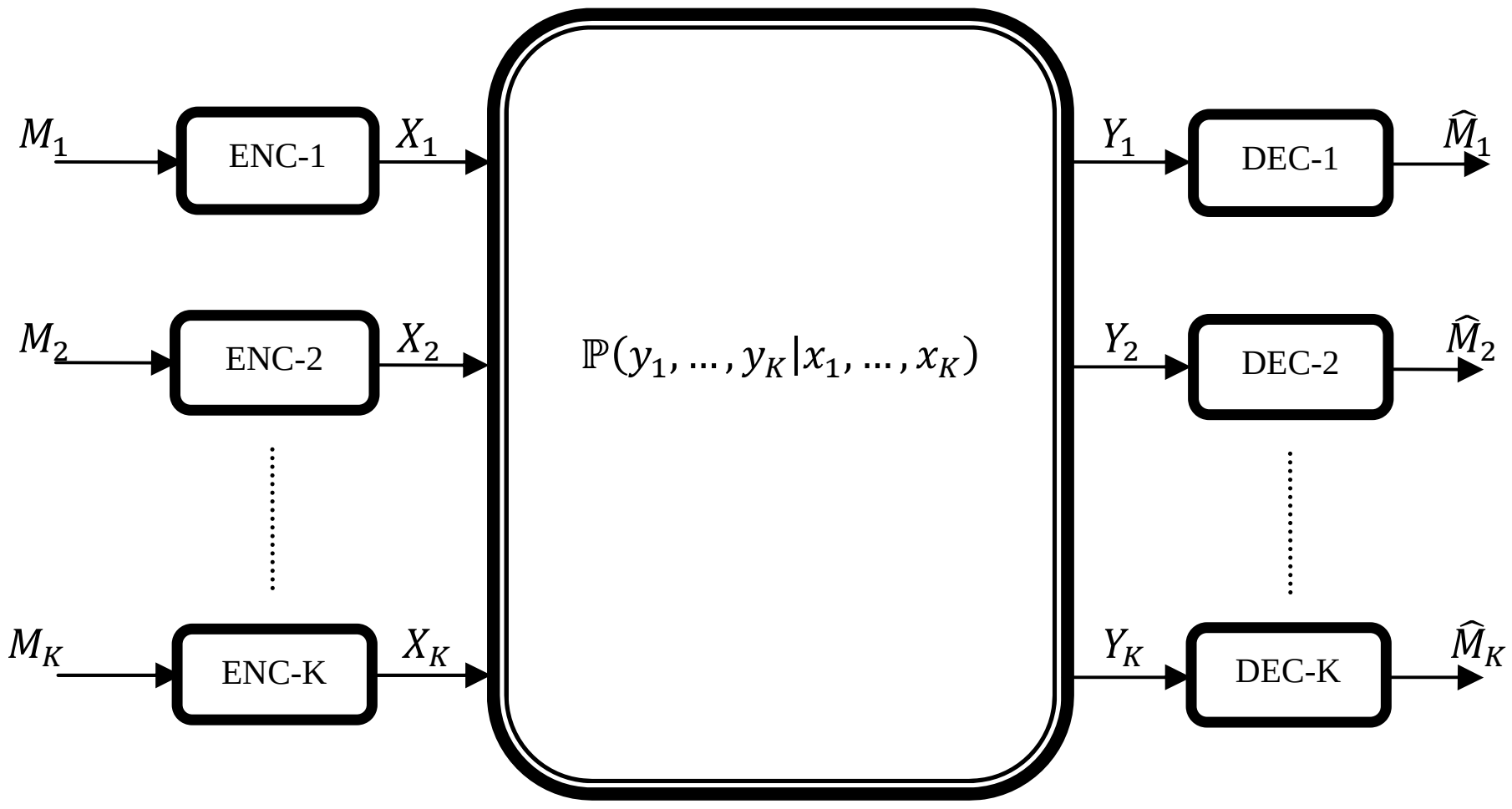

Figure 3. The K-user Classical Interference Channel (CIC).

An open problem in network information theory has been to determine a strong interference regime for this network [9, page 6-68]. We now present a solution to this problem. Specifically, by following the same approach as three-user channel, one can derive the following strong interference regime for the K-user CIC:

**A Strong Interference Regime for the K-User Interference Channel**

$$I(X_2,X_3,X_4,\dots,X_K;Y_K|X_1) \leq I(X_2,X_3,X_4,\dots,X_K;Y_1|X_1) \quad \text{for all joint PDFs} \quad P_{X_2X_3X_4\dots X_K}P_{X_1}$$
$$I(X_1,X_3,X_4,\dots,X_K;Y_1|X_2) \leq I(X_1,X_3,X_4,\dots,X_K;Y_2|X_2) \quad \text{for all joint PDFs} \quad P_{X_1X_3X_4\dots X_K}P_{X_2}$$
$$I(X_1,X_2,X_4,\dots,X_K;Y_2|X_3) \leq I(X_1,X_2,X_4,\dots,X_K;Y_3|X_3) \quad \text{for all joint PDFs} \quad P_{X_1X_2X_4\dots X_K}P_{X_3}$$
$$\vdots$$
$$I(X_1,X_2,\dots,X_{K-1};Y_{K-1}|X_K) \leq I(X_1,X_2,\dots,X_{K-1};Y_K|X_K) \quad \text{for all joint PDFs} \quad P_{X_1X_2\dots X_{K-1}}P_{X_K}$$

(48)

Note that the regime (48) is described by $K$ inequalities. In fact, for the K-user CIC by following the same lines as the three-user CIC, one can derive $\big((K-1)!\big)^K$ different strong interference regime. However, among these regimes, $(K-1)!$ ones are more significant which are derived by exchanging the indices $1,2,3,\dots,K$, with $\vartheta(1),\vartheta(2),\dots,\vartheta(K)$, respectively, in (48) where $\vartheta(.)$ is a *cyclic permutation* of the elements of the set $\{1,2,3,\dots,K\}$. It is worth noting that, up to our knowledge, this is the first time where a full characterization of the capacity region is derived for a general multi-user CIC in a non-trivial case (for both discrete and Gaussian channels).

Let us consider the Gaussian network which is formulated below:

$$\begin{cases} Y_1 = X_1 + a_{12}X_2 + a_{13}X_3 + \dots + a_{1K}X_K + Z_1 \\ Y_2 = a_{21}X_1 + X_2 + a_{23}X_3 + \dots + a_{2K}X_K + Z_2 \\ \vdots \\ Y_K = a_{K1}X_1 + a_{K2}X_2 + \dots + a_{KK-1}X_{K-1} + X_K + Z_K \end{cases} \tag{49}$$

where $Z_1,\dots,Z_K$ are zero-mean unit-variance Gaussian noises and $\mathbb{E}[X_i^2] \leq P_i,\ i=1,\dots,K$. Using Lemma 2, one can show that if the following conditions are satisfied:

$$\begin{cases} \frac{a_{K2}}{a_{12}} = \frac{a_{K3}}{a_{13}} = \dots = \frac{a_{KK-1}}{a_{1K-1}} = \frac{1}{a_{1K}} = \alpha_1 \\ \frac{1}{a_{21}} = \frac{a_{13}}{a_{23}} = \frac{a_{14}}{a_{24}} = \dots = \frac{a_{1K}}{a_{2K}} = \alpha_2 \\ \frac{a_{21}}{a_{31}} = \frac{1}{a_{32}} = \frac{a_{24}}{a_{34}} = \dots = \frac{a_{2K}}{a_{3K}} = \alpha_3 \\ \vdots \\ \frac{a_{K-1,1}}{a_{K1}} = \frac{a_{K-1,2}}{a_{K2}} = \frac{a_{K-1,3}}{a_{K3}} = \dots = \frac{1}{a_{KK-1}} = \alpha_K \end{cases} , \quad |\alpha_i| \leq 1, \quad i=1,2,\dots,K \tag{50}$$

then, the strong interference regime (48) holds. According to the conditions (50), the parameters $a_{1K},a_{21},a_{32},\dots,a_{KK-1}$, which no pair of them lies in either a same row or a same column of the gain matrix, are given by arbitrary real numbers greater than one and the other parameters are given in terms of these parameters by specific relations determined in (50). It is clear that $(K-1)!$ other strong interference regimes are derived by exchanging the indices $1,2,3,\dots,K$, with $\vartheta(1),\vartheta(2),\dots,\vartheta(K)$, respectively, in (50) where $\vartheta(.)$ is a cyclic permutation of the elements of the set $\{1,2,3,\dots,K\}$.

Please refer to our multi-part paper for a detailed systematic study of fundamental limits of communications in interference networks [1-8].

## ACKNOWLEDGEMENT

The author would like to thank F. Marvasti whose editing comments improved the language of this work.